%% file: main.tex
\documentclass[12pt]{article}
\usepackage[margin=1in]{geometry}
\usepackage{lmodern}
\usepackage{fixcmex}
\usepackage[T1]{fontenc}
\input{mymacros.tex}

\input{preamble.tex}

\usepackage{blkarray}
\usepackage{booktabs}

\newcommand{\Dr}{D_{\!R}}

\DeclareFontShape{OMX}{cmex}{m}{n}{%
  <-7.5> cmex7
  <7.5-8.5> cmex8
  <8.5-9.5> cmex9
  <9.5-> cmex10
}{}
\DeclareSymbolFont{largesymbols}{OMX}{cmex}{m}{n}

\begin{document}

\titleauthor{Inertia suppresses signatures of activity of active Brownian particles in a harmonic potential}{A.~Arredondo\footnotemark[1], C.~Calavitta\footnotemark[1], M.~Gomez\footnotemark[2] , J.~Mendez-Villanueva\footnotemark[1]\footnotemark[3], W.\,W.~Ahmed\footnotemark[2] \lowercase{\textsc{and}} N.\,D.~Brubaker\footnotemark[1]\footnotemark[4]}

\footnotetext[1]{Department of Mathematics, California State University, Fullerton, CA, 92831}
\footnotetext[2]{Department of Physics, California State University, Fullerton, CA, 92831}
\footnotetext[3]{Department of Mathematics, University of California, Riverside, CA 92521}
\footnotetext[4]{Corresponding author:\texttt{nbrubaker@fullerton.edu}}

\begin{abstract}
 A harmonically trapped active Brownian particle exhibits two types of positional distributions---one has a single peak, the other has a single well---that signify steady-state dynamics with low and high activity, respectively. Adding inertia to the translational motion preserves this strict single peak/well classification of the densities but shifts the dividing boundary between the states in the parameter space. We characterize this shift for the dynamics in one spatial dimension using the static Fokker--Planck equation for the full joint distribution of the state space. We derive local results analytically with a perturbation method for a small rotational velocity and then extend them globally with a numerical approach.

\end{abstract}

\section{Introduction}

An intriguing feature of confined but non-interacting active Brownian particles (ABPs) is their ability to accumulate into high-density groups in locations away from those typically occupied by standard Brownian particles~\cite{enculescu2011active,kaiser2013capturing,vachier2019dynamics,teeffelen2008dynamics}. 
For instance, ABPs do not spread equally throughout a domain enclosed by rigid walls~\cite{elgeti2013wall,lee2013active} but, instead, congregate near the boundary, regardless of the perimeter shape~\cite{fily2014dynamics,fily2015dynamics, vladescu2014filling}. 
These dense regions form because each particle has a directional persistence induced by its active velocity that creates recurring collisions at the walls---analogous to a bouncing ball---causing them to stick near the edges of the enclosure until reorienting. 

In convex single-well potentials, high-density regions also appear in non-standard configurations, but their formation is not guaranteed~\cite{pototsky2012active}. Two types of positional distributions are possible~\cite{solon2015active,takatori2016acoustic}, depending on the balance of the strength of the trap, the speed of the self-propulsion, and the characteristic rotation of the anterior direction~\cite{basu2019longtime,chaudhuri2021active,malakar2020steady}. If the particle's internal rotation is sufficiently rapid or if its self-propulsive speed is sufficiently small, then the equilibrium distribution is unimodal. Away from these regimes the active force becomes significant enough, when pointing outward, to balance with the inward-pointing potential force and create a distribution with a depression at the minimum of the external potential. 

Adding inertia to dynamics of ABPs further complicates the resulting behavior. Numerous studies~\cite{fily2018mechanical,leoni2020surfing,lowen2019active,sandoval2020pressure,scholz2018Inertial,tapia2021trapped} elucidate features, such as a noise-induced drift~\cite{thiffeault2022anisotropic} or an additional steady-state~\cite{dauchot2019dynamics}, that are not present in (or are at least significantly repressed from) the dynamics of the fully overdamped system. Simulations in~\cite{gutierrez2020inertial} suggest that a non-negligible mass amplifies the effects of the trap and causes the positional density of the particle to smooth, which notability reduces the bias toward the border of the trap in the high-activity the case. We explore this conjecture using analytic and numerical methods.

In this article, we characterize the stationary positional dynamics of inertial ABPs, or active Langevin particles~\cite{lowen2020inertial}, in a harmonic potential. For simplicity, we restrict the dynamics to one spatial dimension and assume that the internal axis of the particle rotates diffusively. The corresponding dimensionless model is
\be\label{model:abp}
 \dot{X} = V/\sqrt{\eps}, \quad
 \sqrt{\eps} \, \dot{V} + V/\sqrt{\eps} + X = \alpha \cos{\Phi} + \sqrt{2}\,\dot{W}, \quad
 \dot{\Phi} = \sqrt{2 \beta}\tsp \dot{\Omega},
\ee 
where $X(t)$, $V(t)$ and $\Phi(t)$ are the position, the (rescaled) velocity, and the internal orientation angle of the isolated particle, respectively; also, $\dot{\Omega}(t)$ and $\dot{W}(t)$ are independent, standard Gaussian white noises. System~\eqref{model:abp} arises from rescaling time and space in the active Langevin particle model~\cite{lowen2020inertial} by the characteristic values $\tau = \gamma/k$ and $\ell = \sqrt{D \tau}$ for a given drag coefficient $\gamma$, potential stiffness $k$, and translational diffusion coefficent $D$. Such a rescaling produces the dimensionless numbers
\[
 \alpha = \frac{u_0\tau}{\ell}, \quad
 \beta = \Dr\, \tau, \quad
 \eps = \frac{m/\gamma}{\tau},
\]
where $m$ is the mass of the particle, $u_0$ is the mean speed of the self-propulsive velocity, and $\Dr$ is the rotational diffusion coefficient of the angular activity. Quantities $\alpha$ and $\beta$ directly control the translation and rotational components of the activity, while $\eps$ defines a damping/quality factor that determines the importance of inertia. Without noise or activity, the dynamics of \eqref{model:abp} are overdamped for $0 \leq \eps < 1/4$, critically damped for $\eps = 1/4$, and underdamped for $\eps > 1/4$. 

Classifying the stationary dynamics of $X(t)$ involves ascertaining the shape of its stationary density $p(x)$ for every value of $(\alpha,\beta,\eps)$ in the first octant of $\R^3$. 
Since the position is non-Markovian, $p$ is inextricably linked to the invariant joint density $\rho(x,v,\phi)$ of the full process $(X,V,\Phi)$ via marginalization. That is, 
\be\label{marg_dist_original}
 p(x) = \int_{-\infty}^\infty \int_0^{2\pi}\rho(x,v,\phi) \diff \phi \diff v,
\ee
where $\rho$, as dictated by \eqref{model:abp}, satisfies the equilibrium Fokker--Planck equation 
\be\label{FPeq}
 \frac{1}{\eps}\left(\pdd{\rho}{v} +  \pd{(v\rho)}{v}\right) - \frac{v}{\sqrt{\eps}} \pd{\rho}{x}  + \frac{(x - \alpha \cos{\phi})}{\sqrt{\eps}}\pd{\rho}{v} + \beta \pdd{\rho}{\phi} = 0,
\ee
over $(x,v,\phi) \in \R^2 \times [0,2\pi)$. Solutions of \eqref{FPeq} are $2\pi$-periodic in $\phi$ and exponentially decay at infinity in both $x$ and $v$~\cite{pavliotis2014stochastic}. While the resulting problem is linear, its corresponding differential operator is not separable (when $\alpha > 0$), which means that $\rho$ is inherently entangled and Fourier methods will not reduce \eqref{FPeq} to an easily solvable system of decoupled, linear algebraic equations~\cite{risken1996solutions}. Finding $\rho$ and, hence, $p$ requires an alternate approach.

In the next section, we analyze the shape of $p(x)$ by solving \eqref{FPeq} with an asymptotic expansion for $\beta \to 0^+$, which connotes that direction of self-propulsion undergoes minimal rotation. In addition to being physically relevant for many active systems, this limit facilitates the calculation of a leading order solution---and further corrections---by conveniently pushing the angular derivatives of the unknown into the nonhomogeneous portion of the higher-order problems. The ensuing density $p(x)$ recovers two equilibrium states. One distribution is unimodal and signifies that the equilibrium dynamics are qualitatively similar to those of a passive particle in a trap. The other is bimodal, which means that activity is a dominant feature of the dynamics. In the parameter space $(\alpha,\beta,\eps)$, these distributions appear in two distinct regions separated by a smooth surface. When $\beta \ll 1$, this dividing surface has a non-parametric representation $\alpha = \alpha^*(\beta;\eps)$, and our perturbation method produces a local expression for the function $\alpha^*$.

In section~\ref{num_results}, we expand these results globally, away from limiting asymptotic regime, with numerical spectral methods. Since the limit $\eps \to 0^+$ is singular, we construction separate methods for the $\eps = 0$ problem, which has two independent variables since the velocity gets marginalized out of dynamic equations, and the $\eps > 0$ problem, which retains the original three independent variables. In the last section, we contextualize the results, discuss the limitations of the model and propose new avenues of research.

\section{Perturbation theory}

In model~\eqref{model:abp}, let's assume that $\beta \ll 1$. This restriction implies that the rotational component of the particle's self-propulsion is exceedingly slow. Accordingly, we expand the joint density as
\be\label{rho_expansion}
 \rho = \rho_0 + \beta \rho_1 + \beta^2 \rho_2 + \cdots,
\ee
and insert it into \eqref{FPeq}, which generates a sequence of problems for the functions $\rho_i(x,v,\phi)$ that are $2\pi$-periodic in $\phi$ and that  decay exponentially in $x$ and $v$:
\be\label{FPeq_ord0}
 \frac{1}{\eps}\left(\pdd{\rho_0}{v} +  \pd{(v\rho_0)}{v}\right) - \frac{v}{\sqrt{\eps}} \pd{\rho_0}{x}  + \frac{(x - \alpha \cos{\phi})}{\sqrt{\eps}}\pd{\rho_0}{v} = 0
\ee
and
\be\label{FPeq_ordi}
 \frac{1}{\eps}\left(\pdd{\rho_i}{v} +  \pd{(v\rho_i)}{v}\right) - \frac{v}{\sqrt{\eps}} \pd{\rho_i}{x}  + \frac{(x - \alpha \cos{\phi})}{\sqrt{\eps}}\pd{\rho_i}{v} = - \pdd{\rho_{i-1}}{\phi}
\ee
for $i \in \N^+$. 

Equation~\eqref{FPeq_ord0} governs the invariant joint density, $\rho_0$, in the total absence of rotational self-propulsion, i.e., when $\beta = 0$. In this regime, $\Phi$ reduces to a random parameter in the dynamics, and the corresponding active force $\alpha \cos\Phi$ is equivalent to a randomly-directed external force of constant magnitude. The translation motion then contains two applied forces, one from this activity and one from the trap, that conveniently combine into an single effective force that is the gradient of the potential $( X - \alpha \cos\Phi)^2/2$. These new dynamics are analogous to those induced a passive particle in a harmonic trap and, consequently, have a unique invariant density~(cf.~\cite[\S6.1]{pavliotis2014stochastic})
\be\label{rho0}
 \rho_0(x,v,\phi) = \frac{1}{Z_0(\phi)}e^{-H_0(x,v,\phi)}.
\ee
In other words, $\rho_0$ is a Gibbs distribution with partition function $Z_0$ and rescaled Hamiltonian function 
\be\label{H0}
H_0(x,v,\phi)  = \frac{v^2}{2} + \frac{1}{2}(x - \alpha \cos\phi)^2.
\ee

Expression~\eqref{rho0} gives the general solution of~\eqref{FPeq_ord0}. In it, the function $Z_0(\Phi)$ is arbitrary and remains unresolved, due to the singular nature of the perturbation, without information about the dynamics for $\beta \neq 0$. We introduce necessary information by requiring continuity between \eqref{rho0} and the solution of full problem~\eqref{FPeq} in the limit $\beta \to 0^+$. If $\beta = 0$, then the angular dynamics reduce to $\dot{\Phi} = 0$, and the long-time marginal distribution of $\Phi$ in \eqref{model:abp} is equivalent to the posited initial distribution. However, if $\beta > 0$, then $\Phi$ diffuses to a uniform distribution over $[0,2\pi)$, regardless of its initial condition. These two cases match only if the initial angle $\Phi(0)$ is uniformly distributed over $[0,2\pi]$. And making such a choice allows us to calculate $Z_0$ from the solution of \eqref{FPeq_ord0} with $\beta = 0$. By integrating \eqref{model:abp}, we have that 
\[
 X(t) = \alpha \, a_1(t)\cos{U} + N_1(t), \quad 
 V(t) = \alpha \, a_2(t) \cos{U} + N_2(t), \quad
 \Phi(t) = U,
\]
where $U$, $N_1$ and $N_2$ are random variables and $a_1$ and $a_2$ are deterministic functions satisfying $a_1(\infty) = 1 $ and $a_2(\infty) = 0$. Further, $U$ is uniformly distribution in $[-\pi,\pi]$ and independent of the random vector $(N_1,N_2)$, which more specifically is a multivariate normal whose components become independent, standard Gaussians themselves when $t \to \infty$. A standard change of variables implies that the stationary joint density of $(X,V,\Phi)$ equals $f_U(\phi)f_{N_1}(x - \alpha \cos{\phi})f_{N_2}(v)$. Comparing this expression with  equation \eqref{rho0} shows that the partition function
\be\label{Z0}
 Z_0 = 4 \pi^2.
\ee 

Next, to find the first order correction $\rho_1$, we introduce the substitution 
\be\label{rho1_sol}
 \rho_1 = \rho_0 \eta_1.
\ee
The original function, $\rho_1$, is in the weighted Hilbert space $L^2(\R^2;  e^{H_0(x,v,\phi)} \diff x \diff v)$---a restriction enforced by its governing differential operator. Hence, the new function $\eta_1$ must be an element of the re-weighted Hilbert space $L^2(\R^2,e^{-H_0(x,v,\phi)}\diff x\diff v)$, which contains a standard basis built from products of Hermite polynomials $\textrm{He}_n$: 
\be\label{hermite_basis}
 \{ \textrm{He}_n(v) \textrm{He}_k(x-\alpha  \cos\phi) \}_{(n,k)\in \N^2}.
\ee 
In this basis,
\be\label{eta1_expn}
 \eta_1(x,v,\phi) = \sum_{(n,k) \in \N^2} c_{n,k}(\phi)\textrm{He}_n(v) \textrm{He}_k(x-\alpha  \cos\phi)
\ee
where $c_{n,k}(\phi)$ is a double sequence of unknown coefficients. 

With \eqref{rho1_sol} and \eqref{eta1_expn}, computing $\rho_1$ becomes a straightforward task. First,~\eqref{rho1_sol} cancels out the exponential factor in~\eqref{FPeq_ordi} (for $i = 1$) induced by $\rho_0$ and transforms the non-homogeneous term of the new partial differential equation for $\eta_1$ to a polynomial in $x$ and $v$. Then \eqref{eta1_expn} reduces this equation to an identity between bivariate power series, which upon equating coefficients produces a finite sum for $\eta_1$: 
\be\label{eta1_form}
 \bead
 \eta_1 
 & = - \alpha \cos \phi \, \textrm{He}_1(x-\alpha \cos\phi)
 + \alpha \stsp\sqrt{\eps}  \cos \phi \, \textrm{He}_1(v)
 + \frac{\alpha^2\eps}{2}  \sin^2\phi \,\textrm{He}_2(v)\\
 & \qquad
 - \alpha^2 \sqrt{\eps} \sin^2\phi \,\textrm{He}_1(v)\textrm{He}_1(x-\alpha \cos\phi) 
 + \frac{\alpha^2 (1 + \eps)}{2}  \sin^2\phi \, \textrm{He}_2(x-\alpha \cos\phi).
 \eead
\ee
While $c_{0,0}(\phi)$ is untouched in the matching process, its value is set to zero in~\eqref{eta1_form}. We justify this choice with a standard Fredholm solvability condition for $\rho_2$ at the next order: problem~\eqref{FPeq_ordi}, for $i = 2$, has a solution only if
\be\label{solv_cond1}
 \iint_{\R^2}\zeta \pdd{\rho_{1}}{\phi} e^{H_0(x,v,\phi)} \diff x \diff v = 0
\ee
for any function $\zeta \in L^2(\R^2; e^{H_0(x,v,\phi)} \diff x \diff v)$ satisfying
\[
 \frac{1}{\eps}\left(\pdd{\zeta}{v} +  \pd{(v\zeta)}{v}\right) + \frac{v}{\sqrt{\eps}} \pd{\zeta}{x} - \frac{(x - \alpha \cos{\phi})}{\sqrt{\eps}}\pd{\zeta}{v}
 = 0.
\]
Solutions of this homogenenous, linear partial differential equation (which may be constructed in the same manner as $\rho_1$) are of the form $\zeta = a(\phi) e^{-H_0(x,v,\phi)}$. Thus, \eqref{solv_cond1} reduces to
\be\label{int_condition_solv}
 0 = \iint_{\R^2} \pdd{\rho_{1}}{\phi} \diff x \diff v.
\ee
Direct integration reduces \eqref{int_condition_solv} to $\ssp{c_{0,0}}''(\phi ) = 0$, i.e., $c_{0,0}(\phi) = C_0 + C_1 \, \phi$. By requiring $2\pi$-periodicity and by making $\int_0^{2\pi}\iint_{\R^2} \rho_1 \diff x \diff v\diff \phi = 0$, which ensures that the total probability of $\rho$ is $1$, we deduce that $C_0 = C_1 = 0$; hence, $c_{0,0}(\phi) \equiv 0$. Plugging \eqref{eta1_form} into \eqref{rho1_sol} produces the full first-order correction $\rho_1$.

A similar process for obtaining $\rho_1$ also generates an exact expression for the second-order correction~$\rho_2$. In other words, we set 
\be\label{rho2_sol}
 \rho_2 = \rho_0 \eta_2,
\ee
then expand $\eta_2$ in the Hermite basis in \eqref{hermite_basis}, and finally fix the coefficients with matching and with the Fredholm solvability condition for the order $\bigoh(\beta^3)$ problem. These steps imply that
\be\label{eta2_form}
 \bead
 \eta_2 
 & = \alpha (1- \eps ) \cos{\phi}\, \text{He}_1(x - \alpha \cos{\phi}) - \frac{\alpha^2(1 - (4\eps^2 + 5) \cos{2 \phi})}{4} \, \text{He}_2(x - \alpha \cos{\phi}) \\
 & \qquad -\frac{\alpha^3 (23\eps^2+ 21\eps + 14)\sin^2{\phi} \,\cos{\phi}}{6 (2 + \eps)} \, \text{He}_3(x - \alpha \cos{\phi}) \\
 & \qquad 
   + \frac{(1 + \eps)^2\alpha^4 \sin^4{\phi}}{8}  \, \text{He}_4(x - \alpha \cos{\phi}) + \xi_2(x,v,\phi),
 \eead
\ee
The function $\xi_2(x,v,\phi)$, which is known but not explicitly written, contains the additive terms with Hermite polynomials of $v$ and, notably, vanishes when integrated with respect to $v$ over $\R$ with the exponential weight $e^{-v^2/2}$.

Accordingly, the three-term expansion of the joint distribution for $\beta \ll 1$ becomes 
\be\label{rho_expansion_2}
 \rho(x,v,\phi) = \frac{\sqrt{\eps}}{4 \pi^2}e^{-H_0(x,v,\phi)} \left(  1 + \beta \,\eta_1(x,v,\phi) + \beta^2 \,\eta_2(x,v,\phi)  + \cdots \right),
\ee
with $\eta_1$ and $\eta_2$ defined in \eqref{eta1_form} and \eqref{eta2_form}. 

Computing the stationary positional density $p(x)$ from \eqref{rho_expansion_2} entails marginalizing out the $v$ and $\phi$ variables. Integrating \eqref{rho_expansion_2} with respect to $v$ is straightforward. All the functions of $v$ multiplicatively separate from those involving $x$ and $\phi$, and the explicit computation reduces to evaluating integrals of Hermite polynomials with exponential weights $e^{-v^2/2}$. Since these polynomials are orthogonal in $L^2(\R,e^{-v^2/2}\diff v)$, each expression with a Hermite polynomial in $v$ of positive degree becomes zero. Accordingly, 
\be\label{rho_margin_wrt_v}
 \int_{-\infty}^\infty \rho(x,v,\phi) \diff v = \frac{e^{-(x -\alpha\cos\phi)^2/2}}{2 \sqrt{2}\pi^{3/2}}  \big( 1 -  \beta g_1(x,\phi) + \beta^2 g_2(x,\phi) + \bigoh(\beta^2)   \big),  
\ee
where $g_1$ and $g_2$ are the functions
\[
 g_1(x,\phi) = \alpha \cos\phi \, (x - \alpha\cos\phi) -  \frac{1 + \eps}{2} \alpha^2  \sin^2\phi\, ((x - \alpha\cos\phi)^2 - 1),
\]
\[
 \bead
 g_2(x,\phi) & = \alpha (1- \eps ) \cos{\phi}\, \text{He}_1(x - \alpha \cos{\phi}) - \frac{\alpha^2(1 - (4\eps^2 + 5) \cos{2 \phi})}{4} \, \text{He}_2(x - \alpha \cos{\phi}) \\
 & \qquad -\frac{\alpha^3 (23\eps^2+ 21\eps + 14)\sin^2{\phi} \,\cos{\phi}}{6 (2 + \eps)} \, \text{He}_3(x - \alpha \cos{\phi}) \\
 & \qquad 
   + \frac{(1 + \eps)^2\alpha^4 \sin^4{\phi}}{8}  \, \text{He}_4(x - \alpha \cos{\phi}).
 \eead
\]

Integrating~\eqref{rho_margin_wrt_v} with respect to $\phi$ is difficult. No explicit antiderivative of the integrand exists; however, the unevaluated result simplifies appreciably with the observation that
\[
 g_i(x,\phi) 
  = h_i(x,\phi) + e^{(x -\alpha\cos\phi)^2/2}\pd{}{\phi}\left(\sin\phi \, G_i(x,\phi)  e^{-(x -\alpha\cos\phi)^2/2} \right), \quad 
\]
for $i =1,2$ and the two functions  
\be\label{U1eq}
 h_1(x,\phi) = \frac{1-\eps}{2} \alpha \cos\phi \, (x - \alpha\cos\phi) 
\ee
\be\label{U2eq}
 \bead
 h_2(x,\phi) & = - \frac{\alpha  (1 + 2 \eps)^2}{4} \cos\phi \, (x - \alpha\cos\phi)  - \frac{\alpha^2}{4}(1 - (5 + 4 \eps^2) \cos^2{\phi})\,\text{He}_2(x - \alpha \cos\phi) \\
 & \qquad + \frac{\alpha^3 (9 \eps^3 - 56 \eps^2 - 39 \eps - 38) \sin^2\phi \cos \phi }{24 (2 + \eps)}\,\text{He}_3(x - \alpha \cos\phi).
 \eead
\ee
In these identities, $G_1$ and $G_2$ are specific bivariate polynomials of $x$ and of sines and cosines of $\phi$. Also, the expressions inside the derivatives vanish at $\phi = 0$ and $\phi = 2\pi$. Only the terms involving $h_1$ and $h_2$ remain after dividing through by the exponential $e^{(x -\alpha\cos\phi)^2/2}$ and integrating.

\begin{figure}[htbp]
\begin{center}
\includegraphics[width=1\textwidth]{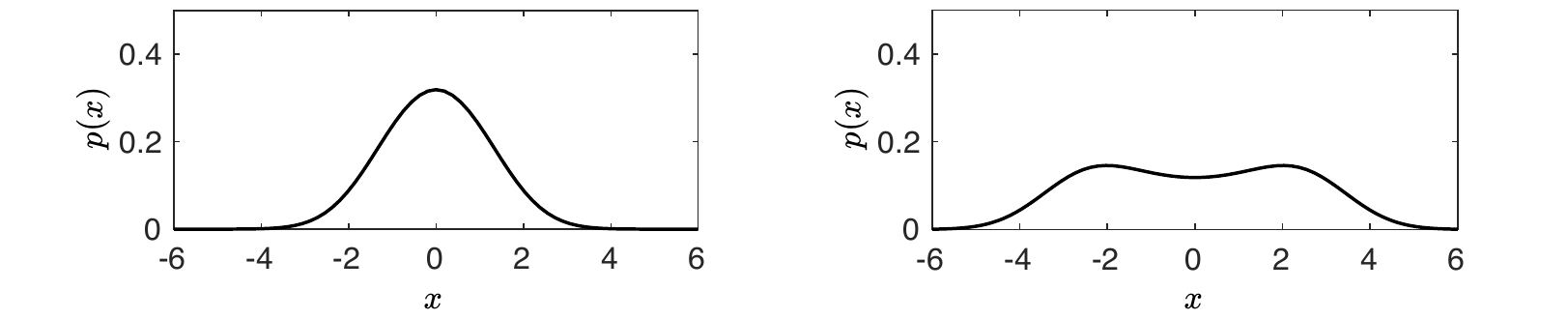}
\caption{Graphs of $p(x)$ from \eqref{p_expn} for $(\alpha,\beta,\eps) = (1,0.1,0.5) $ (left) and $(\alpha,\beta,\eps) = (3,0.1,0.5) $ (right).} 
\label{p_asym}
\end{center}
\end{figure}

As a result, the expansion of the positional density $p(x)$ in the limit $\beta \to 0^+$ is 
\be\label{p_expn}
 p(x) =  \int_0^{2\pi}\frac{e^{-(x -\alpha\cos\phi)^2/2}}{2 \sqrt{2}\pi^{3/2}} \big( 1 -  \beta \, h_1(x,\phi) + \beta^2 h_2(x,\phi) +  \bigoh(\beta^3) \big) \diff \phi
\ee
for $h_1$ and $h_2$ given in \eqref{U1eq} and \eqref{U2eq}. Figure~\ref{p_asym} displays graphs of $p$ for two different sets of parameters values. These values highlight that the distribution is either unimodal or bimodal.
Because $p(x)$ is even (which is expected since the equations of motion have no left-right bias in the direction of movement), the concavity of the $p$ at the origin is a distinguishing property of these modal states. The transition between them is set by the values of $(\alpha,\beta,\eps)$ where the concavity of $p$ changes sign, i.e., where $p''(0;\alpha,\beta,\eps) = 0$. This condition defines an implicit surface that splits the parameter space. We first calculate this surface for $\beta = 0$.

When $\beta = 0$, \eqref{p_expn} contracts to a leading-order contribution depending only on $\alpha$: 
\be\label{p0}
 p_0(x;\alpha) =  \int_0^{2\pi}\frac{e^{-(x -\alpha\cos\phi)^2/2}}{2 \sqrt{2}\pi^{3/2}} \diff \phi.
\ee
 Without $\beta$ and $\eps$, the splitting surface is the hyperplane $\alpha = \alpha_0^*$, where $\alpha_0^*$ is a zero of $p_0''(0;\alpha)$. Differentiating \eqref{p0} twice, evaluating the result at $x = 0$, and then computing the analytic expression of the integral with respect to $\phi$ implies that 
\[
 p_0''(0;\alpha) = \frac{e^{-\alpha ^2/4}}{2 \sqrt{2 \pi }} \big( (\alpha^2-2) I_0(\alpha^2/4) - \alpha^2 I_1(\alpha^2/4)\big),
\]
where $I_0(\cdot)$ and  $I_1(\cdot)$ are modified Bessel functions of the first kind. From this expression, we deduce that $p_0''(0;\alpha)$ has one positive zero, 
\be\label{alpha0}
 \alpha_0^* = 1.7776\ldots.
\ee

This value marks the separation between the two qualitative shapes of \eqref{p0}. That is, if $\alpha \leq \alpha_0^*$, $p_0$ is unimodal (and limits to a standard Gaussian distribution when $\alpha \to 0^+$). Otherwise, $p_0$ is bimodal. Further, as $\alpha \to \infty$, $p_0(x)$ is asymptotic to an arcsine density supported on $[-\alpha,\alpha]$. (Even symmetry in $\phi$ and the substitution $u = \cos\phi$ transform the formula for $p_0$ to
\[
p_0(x) = \frac{1}{\pi  \alpha } \int_{-1}^1\frac{ e^{- \alpha^2 (u - x/\alpha)^2/2}}{\sqrt{2 \pi} \alpha^{-1}}\frac{1}{\sqrt{1-u^2}} \diff u.
\]
The first portion of the integrand is a normal distribution with mean $-x/\alpha$ and variance $\alpha^{-1}$ and is asymptotic to the delta function $\delta(u -x/\alpha)$ for $\alpha \gg 1$. Making this asymptotic replacement yields that $p_0(x) = 1/(\pi  \sqrt{\alpha ^2-x^2})$ for $|x| < \alpha$ and zero otherwise.)

To calculate the higher order corrections to $
\alpha = \alpha_0^*$ in the limit $\beta \to 0^+$, we assume that the threshold is a hypersurface, $\alpha = \alpha^*(\beta,\eps)$, and expand it in a regular perturbation:
\be\label{alpha_star0}
 \alpha^*(\beta,\eps) = \alpha_0^* + \beta \alpha_1^*  + \beta^2 \alpha_2^*  + \cdots.
\ee
Inserting this power series into the zero-concavity condition $p''(0;\alpha^*,\beta,\eps) = 0$ generates two algebraic problems for $\alpha_1^*$ and $\alpha_2^*$ at orders $\bigoh(\beta)$ and $\bigoh(\beta^2)$, respectively. These problems reduce to linear equations after dividing out the nonzero terms and simplifying the resulting expressions with the identity $(\alpha^2-2) I_0(\alpha ^2/4) = \alpha ^2 I_1(\alpha ^2/4)$ for $\alpha =\alpha_0^*$. Their solutions are 
\be\label{alpha_star12}
 \alpha_1^* = \alpha_0^*\frac{ (1 - \eps)}{2}, \quad
  \alpha_2^* =  \alpha_0^* \frac{3 (10 {\alpha_0^*}^2 + 23) \eps^3 + 8 (1-{\alpha_0^*}^2) \eps ^2+( 1 + 2 {\alpha_0^*}^2) (2-3 \eps )}{24 (2+ \eps)}.
\ee

Unfortunately, as $\eps \to \infty$, diverging terms appear in these expressions for $\alpha_1^*$ and $\alpha_2^*$, and ruin the asymptotic hierarchy of \eqref{alpha_star0} when $\eps = \bigoh(\beta^{-1})$ for $\beta \ll 1$. Correcting the divergence requires renormalization~\cite{chen1996renormalization,kirkinis2008secular}. First, we isolate the singular sum $\alpha_S^*$ of \eqref{alpha_star0}, i.e., all the additive terms that grow without bound as $\eps$ increases and induce asymptotic disordering. By factoring out the common components, we deduce that
\[
 \alpha_S^* = -\frac{\alpha_0^*}{2}\eps\beta\left( 1 - \frac{3 (10 {\alpha_0^*}^2 + 23) \eps + 8 (1-{\alpha_0^*}^2)  }{ 12 (2+ \eps)} \eps \beta  + \cdots\right).
\]
The parenthetical expression is asymptotic to the two-term Taylor expansion of the rational function $z/(1 + c(\eps) z)$ at $z = 0$ with $z = \eps \beta$; hence, 
\be\label{alphaS}
 \alpha_S^* \sim -\frac{\alpha_0^*}{2} \frac{\eps\beta}{1 + c(\eps) \eps \beta + \cdots},
  \qquad
 c(\eps) = \frac{3 (23 + 10 {\alpha_0^*}^2 ) \eps + 8 (1-{\alpha_0^*}^2)  }{ 12 (2+ \eps)},
\ee
where $c(\eps)$ is finite for all $\eps \geq 0$. This new expression remains bounded as $\eps \to \infty$, removing the divergence induced by the original form of $\alpha_S^*$. 

The renormalized expression in \eqref{alphaS} of the divergent sum then transforms the threshold expansion in \eqref{alpha_star0} to
\be\label{alpha_star1}
 \alpha = \alpha^*(\beta,\eps) = \alpha_0^* \left( 1 + \frac{\beta }{2} + \frac{(1 + 2 {\alpha_0^*}^2)  (2-3 \varepsilon )}{24 (2 + \eps)} \beta^2 + \cdots\right)  -\frac{\alpha_0^*}{2} \frac{\eps\beta}{1 + c(\eps) \eps \beta + \cdots},
\ee
for $\beta \ll 1$. This local expression produces a surface $(\alpha^*(\beta,\eps),\beta,\eps)$ that splits parameter space into points that yield either bimodal or unimodal distributions. In particular, if $\alpha > \alpha^*(\beta,\eps)$, the equilibrium positional distribution is bimodal. Otherwise, the positional distribution has a single mode at the center of the trap, and the limiting dynamics induced by \eqref{model:abp} are qualitatively similar to those exhibited by a passive Brownian particle. In both cases, $\beta$ must be sufficiently small so that $\eqref{alpha_star1}$ remains valid. 

\begin{figure}[!b]
\begin{center}
\includegraphics[width=1\textwidth]{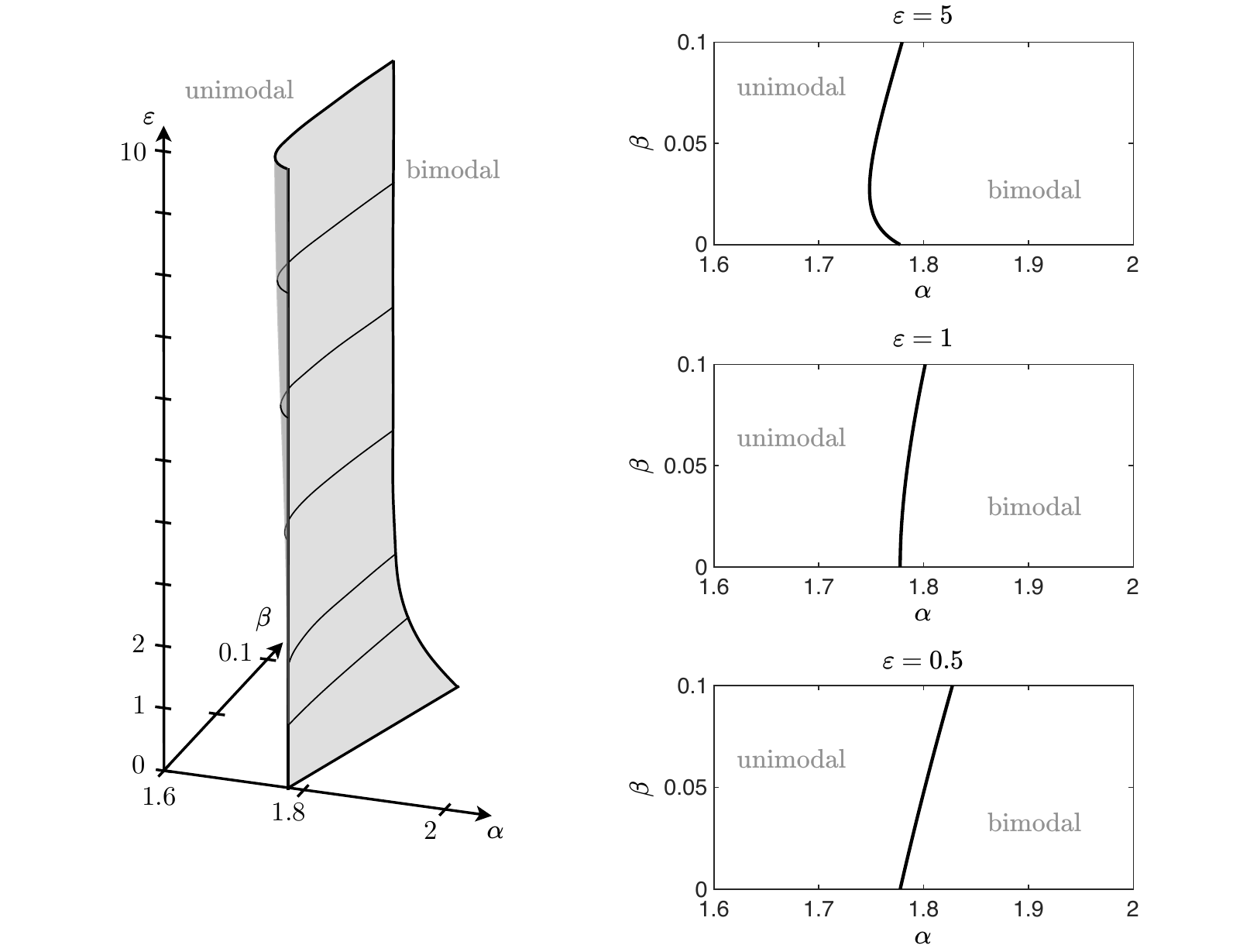}
\caption{Left: Plot of the surface \eqref{alpha_star1} in the parameter space $(\alpha,\beta,\eps)$. Right: Horizontal slices of the surface $\alpha = \alpha^*(\beta,\eps)$ for $\eps = 0.5, 1$ and $5$ (bottom to top).} 
\label{threshold_asym}
\end{center}
\end{figure}

Figure~\ref{threshold_asym}(left) gives a local plot of \eqref{alpha_star1}. A natural way to described its structure is to take horizontal slices for fixed values of $\eps$ and investigate how the resulting curves $(\alpha^*(\beta,\eps),\beta)$ vary as $\eps$ increases, i.e. as inertia becomes more important in the translational dynamics. Figure~\ref{threshold_asym}(right) gives a sequence of such plots. In each slice, the curves emanate from the same point, $(\alpha_0^*,0)$, on the $\alpha$-axis, however, how they enter the space changes. For $\eps \in [0,1)$, $\alpha^*$ initially increases; for $\eps \in [1,\infty)$, $\alpha^*$ decreases; and at $\eps = 1$, $\alpha^*$ initially remains fixed (to first order). Consequently, the bimodal-distribution region near the line $\beta = 0$ expands as $\eps$ becomes larger.

\section{Numerical results}\label{num_results}

Given that \eqref{alpha_star1} is only a local view of the unimodal--bimodal boundary, we next extend the dividing surface into the full parameter space using numerical methods. The approach reverses the steps used to created Figure~\ref{threshold_asym}. First, we fix $\eps$ and compute the curve that partitions the parameter space $(\alpha,\beta)$ into regions that generate either unimodal or bimodal positional distributions. Afterwards, we alter $\eps$ and recompute the curve to study the changes. 

For simplicity, we start at $\eps = 0$. Although this limit is singular, standard perturbation techniques for averaging over the velocity variable in \eqref{FPeq} (see~\cite{doering1990modeling}) reduce the unknown joint density $\rho$ to a function $r(x,\phi)$ that solves 
\be\label{r_PDE}
 \pdd{r}{x} +  \pd{((x - \alpha \cos{\phi})r)}{x} + \beta \pdd{r}{\phi} = 0
\ee
over the domain $\R \times [0,2\pi)$. In this infinite strip, $r$ has periodic boundary conditions in $\phi$ and an exponentially decaying far field behavior in $x$. Also, $r$ satisfies the conservation of probability equation
\be\label{marg_dist}
 \int_{-\infty}^\infty\int_0^{2\pi}r(x,\phi) \diff \phi \diff x = 1.
\ee 
As before, we identify the values of $\alpha$ and $\beta$ at the unimodal--bimodal boundary with the requirement that the marginalized positional distribution has zero-concavity at $x = 0$: 
\be\label{p_xx_eq_0}
  \int_0^{2\pi}\pdd{r}{x}(0,\phi) \diff \phi =  0,
\ee 
System~\eqref{r_PDE}--\eqref{p_xx_eq_0} has three equations and three unknowns $(r,\alpha,\beta)$; however, \eqref{marg_dist} fixes a scaling symmetry of $r$ present in the partial differential equation. So there is essentially one less equation than unknown and, thus, a one-parameter family of solutions to~\eqref{r_PDE}--\eqref{p_xx_eq_0}. Appendix~\ref{app_overdamped} outlines how to numerically trace this family of solutions.

\begin{figure}[!hbt]
\begin{center}
\includegraphics[width=1\textwidth]{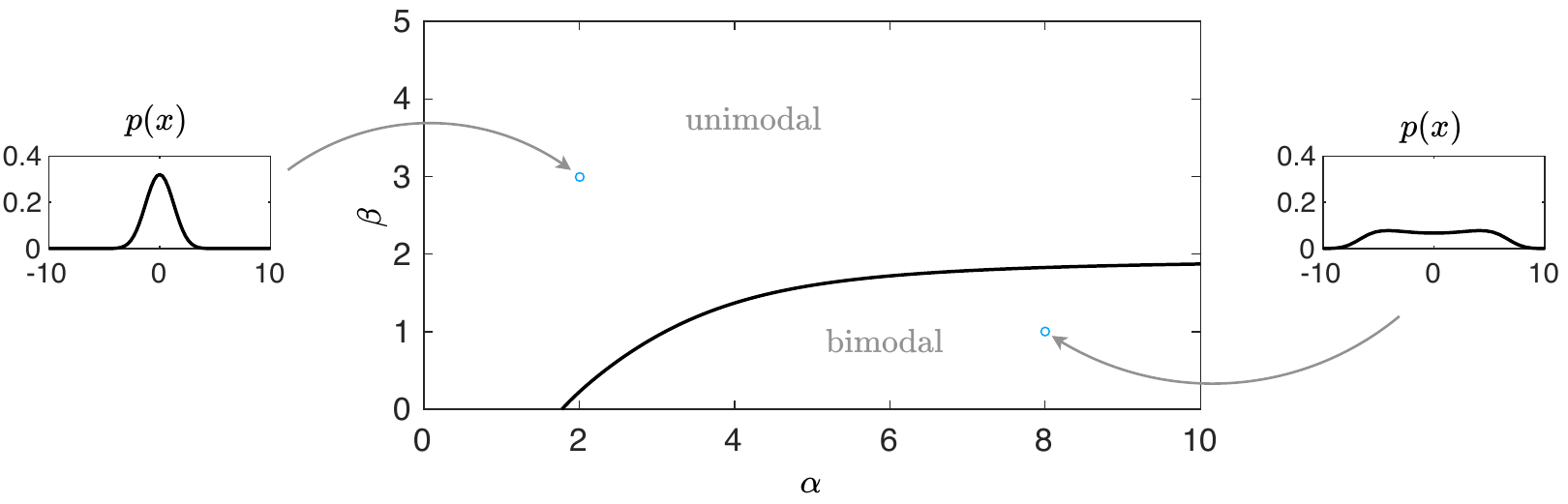}
\caption{For $\eps = 0$, division of the parameter space $(\alpha,\beta)$ into regions whose values generate unimodal and bimodal equilibrium densities. The insets show representative positional densities $p(x)$ in each region.} 
\label{threshold_eps0}
\end{center}
\end{figure}

Figure~\ref{threshold_eps0} shows the computed dividing curve in the $(\alpha,\beta)$-parameter space for $\eps = 0$. Points in the upper region produce positional distributions that have one mode, while those in the lower region produce positional distributions that have two modes. The insets display representative marginal densities
$
p(x) = \int_0^{2\pi} r(x,\phi) \diff \phi
$
for each region.
Although we display only a finite interval of $\alpha$, numerical evidence suggests that the boundary curve remains bounded below the line $\beta = 2$ for all values of $\alpha$. In other words, a significant portion of the parameter space produces unimodal distributions. The parameter $\beta$ gives a ratio of rates controlling the importance of rotational diffusion. So if the characteristic angular movement of the particle's internal axis is rapid enough (i.e., $\beta$ is sufficiently large), then its long-time behavior appears qualitative similar to those exhibited by a passive particles, regardless of the value of $\alpha$.

At nonzero values of $\eps$, finding the threshold becomes more computationally intensive. There is no limiting procedure for reducing \eqref{FPeq} to a single partial differential equation for a function of two independent variables. Instead, we use a spectral method to transform the full problem into an infinite system of coupled partial differential equations~\cite{gottlieb1977spectral} with two independent variables. By setting
\be\label{rho_expan_in_phi}
\rho(x,v,\phi) = \sum_{k = -\infty}^{\infty} u_{k}(x,v) \frac{e^{i k \phi }}{\sqrt{2 \pi }},
\ee
the coefficients $u_{k}$ satisfy a sequence of problems indexed by $k \in \Z$:
\be\label{uk_system}
 \frac{1}{\eps} \left( \pdd{u_{k}}{v} + \pd{(v \tsp u_{k})}{v}  \right) -  \frac{v}{\sqrt{\eps}}\pd{u_{k}}{x} + \frac{x}{\sqrt{\eps}}\pd{u_{k}}{v} - \frac{ \alpha }{2\sqrt{\eps}}  \pd{(u_{k - 1} + u_{k+1})}{v} - \beta  k^2 u_{k} 
  = 0.
\ee
Also, given~\eqref{rho_expan_in_phi}, the conservation of probability and threshold conditions become constraints on $u_0$: 
\be\label{u0_extra_conditions}
 \sqrt{2 \pi}\iint_{\R^2} u_{0}(x,v) \diff x \diff v  - 1 = 0, \qquad 
 \int_{\R}\pdd{u_{0}}{x}(0,v) \diff v = 0.
\ee
System~\eqref{uk_system}--\eqref{u0_extra_conditions} is similar in form to~\eqref{r_PDE}--\eqref{p_xx_eq_0} but with a countable number of unknown functions---the $u_{k}$'s---and two free parameters, $\alpha$ and $\beta$. Analogously, for each fixed $\eps$, it has a one-parameter family of solutions. We trace that family by truncating \eqref{rho_expan_in_phi} for some sufficiently large integer $K$ (e.g., $K \approx 20$) and then deploying a numerical method on the resulting finite system via steps mirroring those of $\eps = 0$ case; see in Appendix~\ref{app_underdamped}.

Figure~\ref{threshold_surf_numerics}(right) displays the threshold curves in $(\alpha,\beta)$-space for $\eps = 1/2$, $1$ and $5$. 
\begin{figure}[!bt]
\begin{center}
\includegraphics[width=1\textwidth]{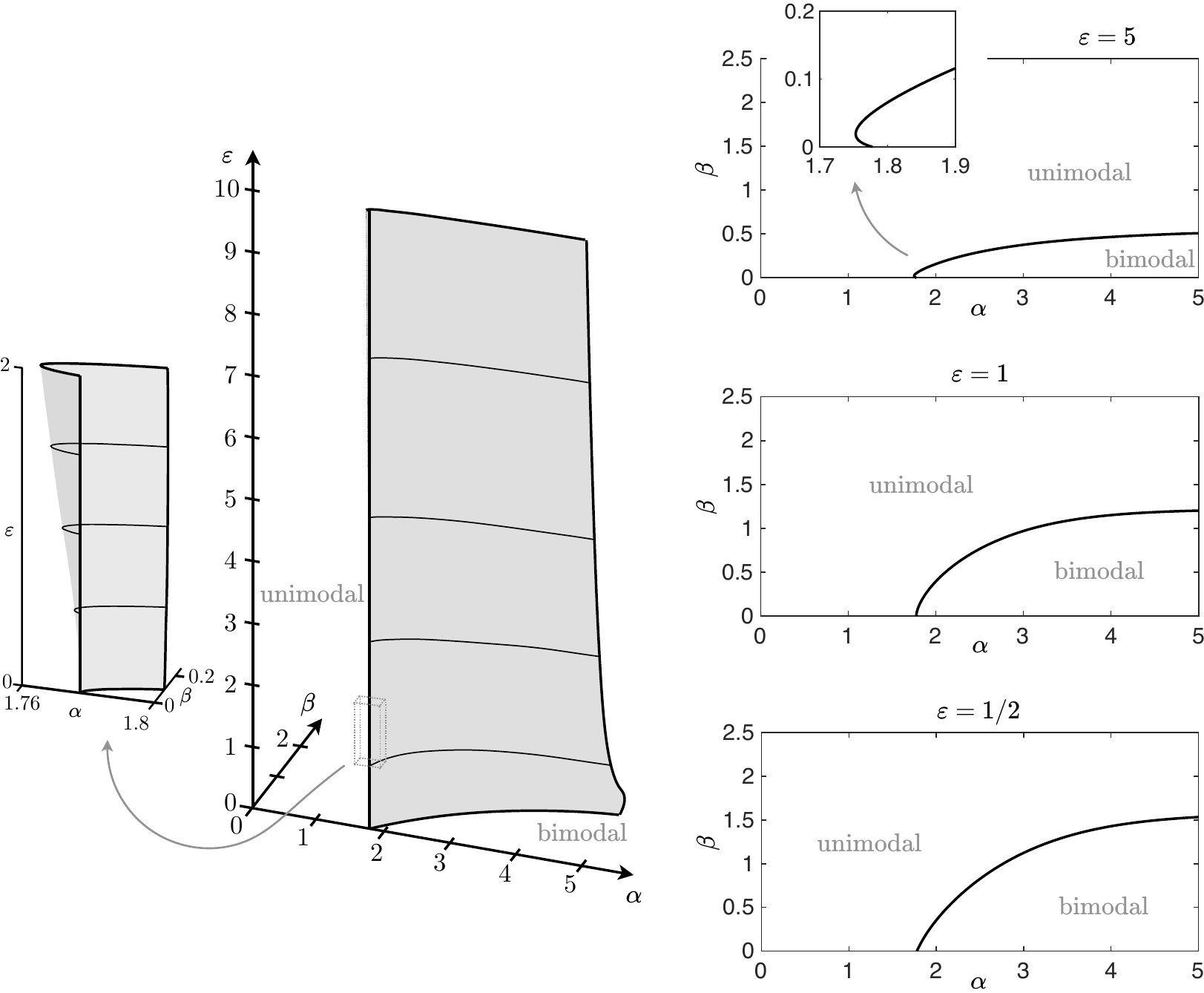}
\caption{Left: Surface splitting the parameter space $(\alpha,\beta,\eps)$ into regions where the harmonically trapped inertial ABP model exhibits of unimodal (left region) and bimodal (right right) equilibrium positional distributions. The inset enlarges a portion of the surface that bulges in the $\alpha$-direction. The bulge appears for $\eps > 1$ and juts out further as $\eps$ increases. Right: Horizontal slices of the surface $\eps = 1/2, 1$ and $5$ (bottom to top). The active (bimodal) domain compresses as $\eps$ grows, which in essence makes the boundary curve fold on itself for $\eps > 1$, as seen in the inset for $\eps = 5$.} 
\label{threshold_surf_numerics}
\end{center}
\end{figure}
Overall, these diagrams are vertically compressed versions of Figure~\ref{threshold_eps0} with the point $(\alpha_0^*,0)$ remaining pinned. Increasing $\eps$ acts as a downward press that displaces the area in Figure~\ref{threshold_eps0} corresponding to bimodal distributions, like squashing a piece of dough with a flat board. Only a small horizontal displacement occurs for $\eps$ in $(0,1]$; however, when $\eps$ becomes greater than $1$, the region bulges over the point $(\alpha_0^*,0)$, causing the dividing curve to fold back on itself. As $\eps$ continues to increase, the bulge juts out farther horizontally, while also thinning vertically. 

Patching the curves together for all values of $\eps$ gives the full threshold surface that divides parameter space $(\alpha,\beta,\eps)$ into points that generate a unimodal positions distributions and points that generate bimodal positional distributions. Figure~\ref{threshold_surf_numerics}(left) gives a plot of the surface, which is an updated, global version of the asymptotic surface in Figure~\ref{threshold_asym}(left). The global surface, as illustrated in Figure~\ref{threshold_surf_numerics}(right) by the curves in its cross sections of constant $\eps$, is more compressed toward zero along the $\beta$-axis than the local surface. For small values of $\eps$, the compression is nominal. But it becomes is especially pronounced for large values of $\eps$. 

\begin{figure}[!htb]
\begin{center}
\includegraphics[width=1\textwidth]{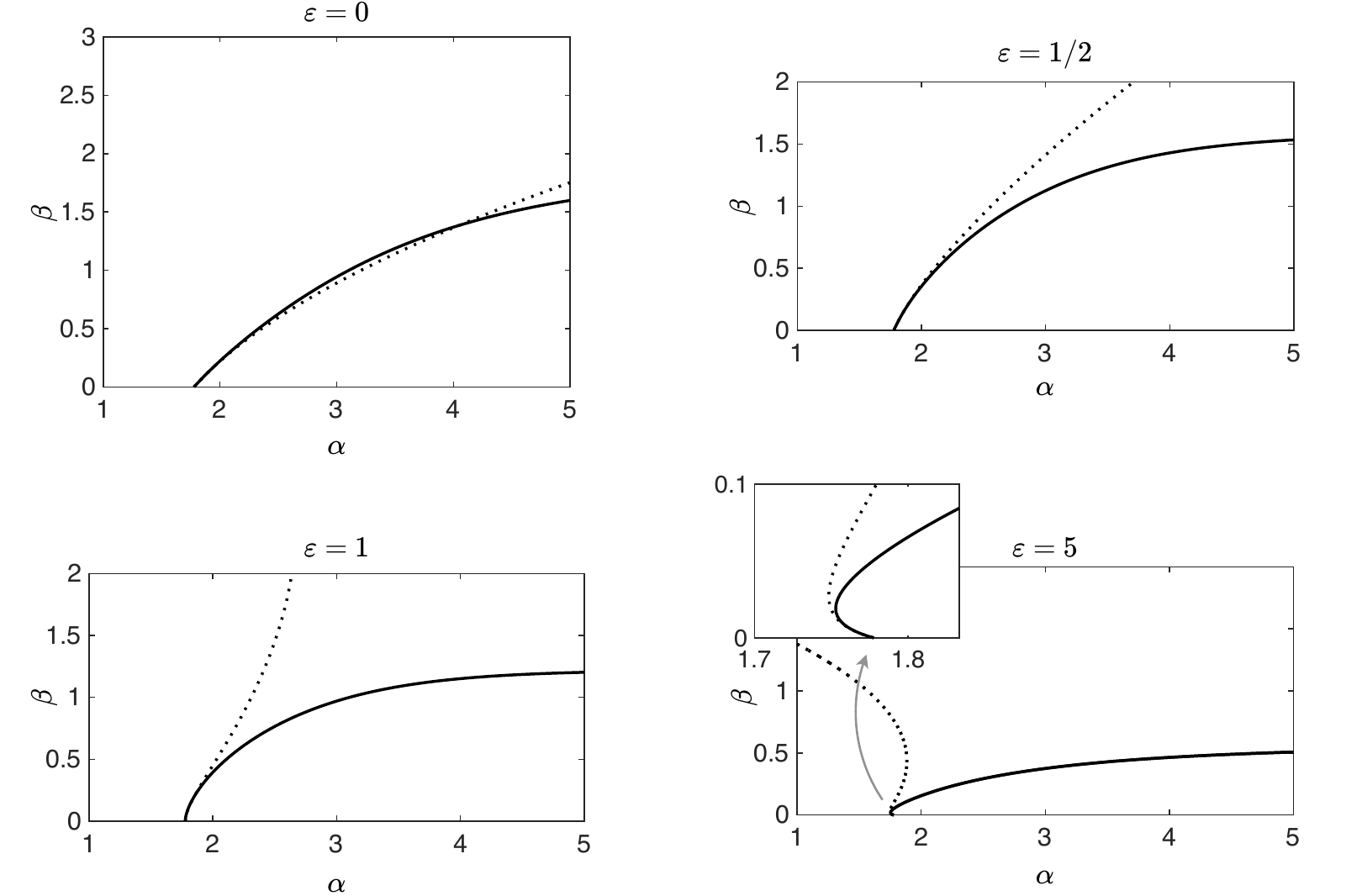}
\caption{Comparison between the asymptotic and numeric calculations (dotted and solid, respectively) of the threshold curve $\alpha = \alpha^*(\beta;\eps)$ between the unimodal and bimodal states for $\eps = 0,1/2,1,$ and $5$. Their is good local agreement between the results, even for $\beta = \bigoh(1)$ in cases, which is outside the region of validity of the asymptotic expansions. For instance, the threshold expansion for $\eps = 1$ remains appropriate out to $\beta = 0.5$.}
\label{threshold_comparison}
\end{center}
\end{figure}

Figure~\ref{threshold_comparison} supplies a comparison of the numerically computed and the asymptotically reconstructed threshold curves from the surfaces's transverse slices for $\eps = 0$, $1/2$, $1$ and $5$. All the expansions perform reasonably well away from the regime $\beta \ll 1$, especially the ones for $\eps = 0$ and $1/2$. These values yield asymptotic curves that give a very good classification out to $\beta = 1$.

\section{Discussion}

The derived results demonstrate that translation inertia substantially influences the behavior of noisy self-propulsive particles. And the changes further obfuscate the non-Gaussian features that clearly distinguish the particles' activity; specifically, adding translational inertia to the overdamped dynamics squashes the parameter space of situations that clearly exhibit activity.

Without inertia (i.e., $\eps = 0$ in \eqref{model:abp}), the characteristic rotational speed, $\beta$, of the self-propulsive axis delineates the dynamics into high and low activity states. If the speed is large (i.e., $\beta \gg 1$), the particle rapidly reorients, meaning there is minimal directional persistence induced by self propulsion, regardless of its speed $\alpha$. So the translational active force becomes analogous in form to thermal noise, which competes with the trap to induce dynamics that are equivalent a Brownian-like particle moving in a harmonic potential. As $\beta$ decreases, the reorientation time extends and eventually creates a non-negligible asymmetry in the movement to and from the center of the trap. When the active force and the trap force align, the particle darts through the center of the potential to a position where they act in opposition (i.e.~the other side of the trap). It remains stuck there while slowly turning around. The process then repeats and, in aggregate, yields a bias toward the boundary where the forces balance and away from the potential's center. But this boundary only appears if the self-propulsive speed $\alpha$ is adequately large. If it isn't, then the active force is too small and the trap dominates. In totality, the parameter regime of strong activity appears to the right of $\alpha = \alpha^*_0 = 1.77761\ldots$ and remains bounded below $\beta = \beta_0^* \approx 2$.

Adding inertia (i.e., having $\eps > 0$) extends the persistence time of the translational motion and, in turn, amplifies the impacts of the trap on the dynamics. Self-propulsive effects are also boosted but less significantly so since the active force randomly changes direction. To generate an asymmetry that skews the positional bias away from the center of the trap, the characteristic reorientation time must slow down, which implies that the values of $\beta$ needed to generate high activity decrease from the previous $\beta = \beta_0^*$ threshold. But the necessary values of characteristic self-propulsive speed $\alpha$ also decrease since inertia, when coupled with nominal rotation, magnifies active force more than the trap. In totality, as $\eps$ grows, the percentage of the parameter space $(\alpha,\beta)$ exhibiting high activity (biomodal distribution) contracts---by shifting down in $\beta$ and slightly expanding $\alpha$; see Figure~\eqref{threshold_surf_numerics}(right). 

While our model involves only one spatial dimension, we expect these qualitative changes to generalize to other situations, including ones with more spatial dimensions, or elliptical/ellipsoidal confining potentials, or both. The physical explanation of the previous paragraph does not depend on the specific geometry, so we anticipate only minor quantitative differences in the dynamics appearing in the varied setups. 

For $\beta \ll 1$, the outlined asymptotic approach remains a powerful exploratory tool for many of these alternate scenarios.  While it may seem restrictive to assume that $\beta$ is small, active systems typically operate in the this regime, where self-propulsion contains a dominant anterior direction that has a slight, irregular rotation induced by a small defect in the drive mechanism. A leading order solution is again an invariant Gibbs distribution of the non-rotational dynamics, whose potential energy contains an extra term accounting for a random initial orientation. Subsequent corrections are power series expansions of Hermite functions, which reduce to finite sums if the trapping force is a polynomial.

Investigating the totality of changes in higher dimensions for all values of $\beta$, however, is difficult. Including inertia in the translation dynamics expands the number of stochastic variables (barring possible symmetry arguments) from three to five in two-space (i.e., $(x,v,\phi)$ to $(x,y,v_1,v_2,\phi)$) and from five to eight in three-space (i.e., $(x,y,v_1,v_2,\phi)$ to $(x,y,z,v_1,v_2,v_3,\phi,\psi)$). The ensuing static Fokker--Planck equation increases in dimension, meaning standard numerical algorithms for approximating its solutions become much more computationally intensive. Monte Carlo methods based on path simulations circumvent this curse of dimensionality, although accurately determining the activity threshold is likely harder. There is no straightforward procedure for plugging these methods into a curve tracing algorithm.

Along with inquiries into the effects of inertia in higher dimensional systems, there are many more problems worth considering. Perhaps the most pressing, regarding the derived results, is determining the scaling law for the unimodal/bimodal threshold as $\alpha \to \infty$. Simulations suggest that $\beta \sim c_0 + c_1 \, e^{-c_2 \alpha}$, which corroborates the work in~\cite{pototsky2012active}, but a formal perturbative approach is not immediately obvious. 

Also, many natural extensions of our basic model exist. For one, our assumed angular dynamics exclude inertia, which implies that particle's internal orientation has no memory of its previous states. Adding angular inertia our model will most likely enhance the percentage of bimodal distributions since it enhances the directional persistence of the translational self-propulsion without affecting the trap. Two, our model uses a simple harmonic potential. Such a choice is common~\cite{dauchot2019dynamics,jahanshahi2017brownian,jones2021stochastic}, however it may be too reductive for exploring the high-low activity threshold. For instance, certain non-harmonic potentials---such as $U(x) = x^4/4$ and $U(x,y) = (x^2+y^2)^2/4$ in one and two dimensions, respectively---yield strong activity for arbitrarily small values of $\alpha$; namely, the anchor point of the threshold curve on the $\beta$-axis is located at $\alpha = 0$~\cite{pototsky2012active}, not at $\alpha = \alpha_0^* >0$. How the threshold enters the parameter space is unknown, with and without inertia. Also, due to the location of anchor point, the highly active regime can not longer bulge in negative $\alpha$-direction for sufficiently large $\eps$. Will the base point remain fixed, or will it slide along the vertical $\beta$-axis at some finite $\eps$? The perturbation method yields a promising approach for answering this question. 

\appendix

\section{Appendix: Numerical Methods}

\subsection{$\eps = 0$}\label{app_overdamped}

Since the solution of \eqref{r_PDE}--\eqref{p_xx_eq_0} decays rapidly decays as $|x| \to \infty$, we pragmatically truncate the infinite spatial domain to a finite interval $[-l,l]$ for a sufficiently large  $l$ and apply homogeneous Dirchlet conditions at $x = \pm l$. A coarse restriction is that $l \gg \max\{\alpha,1\}$, which implies that trapping force is large at $x = l$ and also dominates the self-propulsions (i.e., if $x \gg \alpha \cos\phi$). 

Over this finite domain, we use a pseudospectral method to discretize the problem~\cite{trefethen2000spectral}. That is, we construct a tensor product grid of $[-l,l]\times[0,2\pi]$, with $N+1$ Chebyshev points $x_n = -l \cos(n\pi/N)$ (for $n = 0,1,\ldots,N$) over $[-l,l]$ and $M$ uniformly spaced points $\phi_m = 2\pi(m-1)/M$ (for $m = 1,2,\cdots,M$) extending across $[0,2\pi)$. On the grid we approximate the function $r(x,\phi)$ with a matrix $R$ of unknown values. For our setup, $x$ and $\phi$ vary along the columns and rows, respectively. Also, given that $r$ is a zero at $x = \pm l$, we drop the first and last rows, reducing $R$ to an $M \times (N-1)$ matrix. For the assumed grids and boundary conditions, let $D_{x}^{(k)} \in \R^{(N-1)\times (N-1)}$ and $D_{\phi}^{(k)} \in \R^{M\times M}$ be the $k$-th order pseudospectral differentiation matrices for $x$ and $\phi$. Also, define $\vect{w}_{x} \in \R^{N-1}$ and $\vect{w}_{\phi}\in \R^{M}$ to be column vectors of the Clenshaw--Curtis and trapezodial weights in $x$ and $\phi$. Differentiating and integrating then amounts to right and left matrix multiplication: 
\[
 \begin{gathered}
 \pd{r}{x} \approx R (D_{x}^{(1)})^T, \quad 
 \pd{r}{\phi} \approx D_{\phi}^{(1)} R, \quad
 \pdd{r}{x} r \approx R (D_{x}^{(2)})^T, \quad \cdots \\
 \int_{-l}^{l} r \diff x \approx R \vect{w}_{x}, \qquad
 \int_{0}^{2\pi} r \diff \phi \approx \vect{w}_{\phi}^T R.
 \end{gathered}
\] 
Lastly, we require $N$ to be odd so that zero is a grid point of $x$ and approximating $\partial_{xx}(\cdot)|_{x = 0}$ only involves to extracting the middle row, $\hat{D}_{x}^{(2)}$, of $D_{x}^{(2)}$. 

The system resulting from discretizing equations \eqref{r_PDE}--\eqref{p_xx_eq_0} is
\[
 \begin{gathered}
 R (D_{x}^{(2)})^T + (X - \alpha\, C_\phi) \circ R (D_{x}^{(1)})^T + R + \beta D_{\phi}^{(2)} R = 0, \\
  \vect{w}_{\phi}^T R \tsp \vect{w}_{x} = 1, \qquad 
  \vect{w}_{\phi}^T R (\hat{D}_{x}^{(2)})^T  =  0,
 \end{gathered}
\]
where $X$ and $C_\phi$ are matrices containing the values of $x$ and $\cos\phi$ on the tensor grid, and $\circ$~indicates the pointwise multiplication between the surrounding matrices. Vectorizing each equation (i.e., applying the operator $\mathop{\mathrm{vec}}(\cdot)$) transforms the system to the standard matrix form for $\vect{r} = \mathop{\mathrm{vec}}(R) \in \R^{M(N-1)}$:
\be\label{disc_sys_eps0}
 \begin{gathered}
 (A_1 - \alpha \tsp  A_2 + \beta A_3) \vect{r} = \vect{0}, \\
  W_{\!1} \vect{r} = 1, \qquad 
  W_{\!2} \vect{r}  =  0.
 \end{gathered}
\ee 
In this concise representation, the matrices $A_1$, $A_2$ and $A_3$ and row vectors $
W_1$ and $W_2$ are 
\[
 \begin{gathered}
 A_1 = (D_{x}^{(2)} \otimes I_M)  + \mathop{\mathrm{diag}}(\mathop{\mathrm{vec}}(X))(D_{x}^{(1)} \otimes I_M)  +  I_N\otimes I_M \\
 A_2 =  \mathop{\mathrm{diag}}(\textrm{vec}(C_\phi)) (D_{x}^{(1)} \otimes I_M),
 \qquad 
 A_3 = I_N \otimes D_{\phi}^{(2)},
 \end{gathered}
\]
\[
 W_1 = \vect{w}_{x}^T \otimes \vect{w}_{\phi}^T, \qquad
 W_2 =\hat{D}_{x}^{(2)} \otimes \vect{w}_{\phi}^T,
\]
where $I_N$ and $I_M$ are identities matrices of size $N-1$ and $M$, and $\otimes$ denotes the standard Kronecker product of two matrices.

We compute the family of solutions for \eqref{disc_sys_eps0} in the unknowns $(\vect{r},\alpha,\beta)$ with a continuation method. The initial solution for the algorithm is the threshold solution on the $\alpha$-axis, i.e., the function \eqref{rho_margin_wrt_v} at $(\alpha,\beta) = (1.7776\ldots,0)$, discretized over the tensor grid. To find the next solution, we increment $\beta$ and then solve \eqref{disc_sys_eps0} for $(\vect{r},\alpha)$ with Newton's method, starting from the initial $(\vect{r},\alpha)$-pair at the previous $\beta$. Note that Newton's method is necessary since the equations are nonlinear when $\alpha$ is free. After finding the solution, this process is repeated. From one iteration to the next, we control the step size of $\beta$ to ensure that Newton's method converges and the value of $l$, which must change as $\alpha$ increases to preserve the validity of the trimmed domain. The left panel in Figure~\ref{threshold_eps0} displays the results of the this algorithm. The distributions in the right panel are found from directly computing the solution $\vect{r}$ of \eqref{disc_sys_eps0} for the stated $(\alpha,\beta)$. 

\subsection{$\eps > 0$}\label{app_underdamped}

We also discretize \eqref{uk_system} and \eqref{u0_extra_conditions} with pseudospectral method. The main difference is that now both independent variables sweep out unbounded intervals, meaning that for practical computational purposes the full domain $\R^2$ must be pruned to a rectangle $\Omega_R = [-l_x,l_x] \times [-l_v,l_v]$, for adequately large $l_x$ and $l_v$. At its boundary, each $u_k$ has homogeneous Dirichlet conditions, given that original differential operator in \eqref{FPeq} implies that the solution undergoes rapid exponential decay as $|(x,v)| \to \infty$. Accordingly, we use Chebyshev points in both directions of the tensor grid, since there is no periodicity in $v$, and exclude the boundary points from the computations. Also, to make system~\eqref{rho_expan_in_phi} finite, we introduce a cutoff integer $K$ that removes all the frequencies $k$ such that $|k| \geq K$.

The discrete vectorized system takes a form similar to \eqref{disc_sys_eps0}. Specifically, it is
\be\label{disc_sys_eps}
 \bead
 (\mathcal{A}_1 - \alpha \mathcal{A}_2  - \beta \mathcal{A}_2) \vect{u}  = \vect{0}, \\
  \mathcal{W}_1 \vect{u} = 1, \qquad  \mathcal{W}_2 \vect{u} = 0,
 \eead
\ee
where 
\[
 \vect{u} = 
  \begin{bmatrix}
   \vect{u}_{-K} \\ \vdots \\ \vect{u}_{-1} \\ \vect{u}_{0} \\ \vect{u}_{1} \\  \vdots  \\ \vect{u}_{K}
 \end{bmatrix}, \qquad 
 \mathcal{A}_1 = \begin{bmatrix}
 A_1^\eps &  &  &  &  &  &  0 \\
   & \ddots   \\
   &  & A_1^\eps    \\
   &  &  & A_1^\eps   \\
   &  &  &  & A_1^\eps \\
   &  &  &  &  & \ddots   \\ 
 0 &  &  &  &  &  & A_1^\eps   \\
 \end{bmatrix}
\]
\[
 \mathcal{A}_2 = \begin{bmatrix}
 0 & A_2^\eps &  &  &  &  & 0 \\
 A_2^\eps & 0 & \ddots  \\
   & \ddots & \ddots & A_2^\eps \\
   &  & A_2^\eps & 0 & A_2^\eps  \\
   &  &  & A_2^\eps & \ddots & \ddots  \\
   &  &  &  & \ddots &  0 & A_2^\eps  \\ 
 0 &  &  &  &  & A_2^\eps & 0   \\
 \end{bmatrix}, \qquad
 \mathcal{A}_3 = \begin{bmatrix}
 K^2I &  &  &  &  &  &  0 \\
   & \ddots   \\
   &  & 1^2 I    \\
   &  &  & 0   \\
   &  &  &  & 1^2I \\
   &  &  &  &  & \ddots   \\ 
 0 &  &  &  &  &  & K^2I   \\
 \end{bmatrix}
\]
and 
\[
 \mathcal{W}_1 = \begin{bmatrix}
 0 & \cdots & 0 & W_1^\eps & 0 & \cdots & 0 
 \end{bmatrix}, \qquad
 \mathcal{W}_2 = \begin{bmatrix}
 0 & \cdots & 0 & W_2^\eps & 0 & \cdots & 0 
 \end{bmatrix}.
\]
In these expressions, each $\vect{u}_k$ is the discrete and vectorized version of the coefficient function $u_k(x,v)$ on the grid, and $A_1^\eps$, $A_2^\eps$, $W_1^\eps$ and $W_2^\eps$ are matrix representations of operators that act on those functions; see Table~\ref{operator_tab}. 

Given the equivalent form, we solve \eqref{disc_sys_eps} with the same algorithm as \eqref{disc_sys_eps0}. The only distinctions are a new starting starting point, although $(\alpha,\beta)$ remain the same, and that both $l_x$ and $l_v$ may be modified. Figure~\eqref{threshold_surf_numerics} displays the threshold surface computed from this continuation method.

\begin{table*}[h!]
\centering
\begin{tabular}{cc} \toprule
    Operator & Matrix \\ \midrule
    & \\
    $\vspace{1em}\ds \frac{1}{\eps}\left(\pdd{}{v} +  \pd{}{v}v\right) - \frac{v}{\sqrt{\eps}} \pd{}{x}  + \frac{x}{\sqrt{\eps}}\pd{}{v}$  & $A_1^\eps$  \\
    $\vspace{1em}\ds \frac{ 1 }{2\sqrt{\eps}}\pd{}{v}$  & $A_2^\eps$  \\
    $\vspace{1em}\ds\sqrt{2 \pi}\iint_{\R^2} (\cdot)\diff x \diff v$  & $W_1^\eps$  \\
    $\vspace{1em}\ds\pdd{}{x} \int_{\R} (\cdot) \diff v \big|_{x = 0}$  & $W_2^\eps$  \\ \bottomrule
\end{tabular}
\caption{Operators and the matrices notating their discrete, vectorized versions.}\label{operator_tab}
\end{table*}

\bibliographystyle{siamplain}

{\small \bibliography{main.bib}}

\end{document}

%% file: mymacros.tex
\usepackage{amssymb,amsthm,amsmath,amsopn}
\usepackage[normal]{subfigure}
\usepackage{import}
\usepackage{verbatim}
\usepackage{mathrsfs}
\usepackage{colonequals}
\usepackage{paralist}
\usepackage{ctable}
\usepackage{bm}
\usepackage{mathtools}
\usepackage[colorlinks=true,linkcolor=blue,citecolor=blue]{hyperref}%
\usepackage{microtype}

\graphicspath{{./images/}}

\theoremstyle{definition}

\newcommand{\R}{\mathbb{R}}

\newcommand{\N}{\mathbb{N}}

\newcommand{\Z}{\mathbb{Z}}

\newcommand{\bigoh}{\mathcal{O}}

\newcommand{\be}{\begin{equation}}
\newcommand{\ee}{\end{equation}}
\newcommand{\bea}{\begin{align}}
\newcommand{\eea}{\end{align}}
\newcommand{\bead}{\begin{aligned}}
\newcommand{\eead}{\end{aligned}}
\newcommand{\bsub}{\begin{subequations}}
\newcommand{\esub}{\end{subequations}}
\newcommand{\vect}[1]{\bm{#1}}

\newcommand{\ds}[0]{\displaystyle}

\newcommand{\stsp}{\:\!}
\newcommand{\tsp}{\;\!}
\newcommand{\stnsp}{\!\!\:}
\newcommand{\ssp}[1]{{#1 \stnsp}}

\newcommand*{\diff}{\mathop{}\!{d}}
\newcommand{\pd}[2]{\frac{\partial #1}{\partial #2}}
\newcommand{\pdd}[2]{\frac{\partial^2 #1}{\partial #2^2}}

\newcommand{\eps}{\varepsilon}


%% file: preamble.tex
\usepackage{titlesec}

\newcommand{\titleauthor}[2]{
	\hphantom{.}
	\vspace{.5in}
	\begin{center}
		\textbf{\Large \uppercase{#1}}
	\end{center}
	\begin{center}
		{\small \uppercase{#2}}
	\end{center}}

\titleformat{\title}[hang]{\large\bfseries}{}{0em}{\LARGE\bfseries}[]
\titleformat{\section}[hang]{\bf}{\thesection.}{.5em}{}[]
\titleformat{\subsection}[hang]{\it}{\thesubsection.}{.5em}{}[]
\titleformat{\subsection}[hang]{\it}{\thesubsection.}{.5em}{}[]
\titleformat{\subsubsection}[hang]{\it}{\thesubsubsection.}{.5em}{}[]